\documentclass[epj]{webofc}
\usepackage[varg]{txfonts}   % Web of Conferences font
\usepackage[english]{babel}
\newcommand{\gev}{\,{\rm GeV}}
\wocname{EPJ Web of Conferences}
\woctitle{INPC 2013}
\begin{document}
\title{Next to leading order analysis of DVCS and TCS}
%
% subtitle is optional
\author{J. Wagner\inst{1}\fnsep\thanks{\email{jwagner@fuw.edu.pl}} \and
		H. Moutarde\inst{2} \and
        B. Pire\inst{3} \and
        F. Sabati\'e\inst{2} \and
        L. Szymanowski \inst{1}
        % etc.
}

\institute{National Center for Nuclear Research (NCBJ), Warsaw, Poland
\and
           Irfu-SPhN, CEA, Saclay, France
\and
           CPHT, {\'E}cole Polytechnique, CNRS, 91128 Palaiseau, France
          }

\abstract{%
The study of $O(\alpha_s)$ QCD contributions to the timelike and spacelike virtual Compton scattering  amplitudes in the generalized
Bjorken scaling regime demonstrates that gluonic contributions  are by no means negligible even  in the medium energy range which will be studied intensely at JLab12 and in the COMPASS-II experiment at CERN. }
\maketitle
%%%%%%%%%%%%%%%%%%%%%%%%%%%%%%%%%%%%%%%%%%%%%%%%%%%%%%%%%%%%%%%%%%
\section{Introduction}
%%%%%%%%%%%%%%%%%%%%%%%%%%%%%%%%%%%%%%%%%%%%%%%%%%%%%%%%%%%%%%%%%%
\label{intro}
%%%%%%%%%%%%%%%%%%%%%%%%%%%%%%%%%%%%%%%%%%%%%%%%%%%%%%%%%%%%%%%%%%%%%%%%%%%%%%%%

A necessary step to extract information on  gluon GPDs is to study  \cite{Moutarde:2013qs} $O(\alpha_s)$ QCD contributions to spacelike Deeply Virtual Compton Scattering (DVCS) :
\begin{equation}
\gamma^*(q_{in}) N(P) \to \gamma(q_{out}) N'(P'=P+\Delta) \,,~q_{in}^2 =-Q^2,~q_{out}^2 =0,~t=\Delta^2,~\xi =\frac{Q^2}{(P+P')\cdot(q_{in}+q_{out})}\,,
\end{equation}
and to its crossed reaction, timelike Compton scattering (TCS) :
\begin{equation}
\gamma(q_{in}) N(P)\to \gamma^*(q_{out}) N'(P'=P+\Delta)\,,~q_{in}^2 =0,~q_{out}^2 =Q^2,~t=\Delta^2,~ \eta =\frac{Q^2}{(P+P')\cdot(q_{in}+q_{out})}\,,
\end{equation}
which are the simplest reactions to study generalized parton distributions (GPDs). After factorization, the DVCS (and similarly TCS) amplitudes are written in terms of  Compton form factors  (CFF) $\mathcal{H}$, $\mathcal{E}$ and $\widetilde {\mathcal{H}}$, $\widetilde {\mathcal{E}}$ , as :
\begin{equation}
\mathcal{A}^{\mu\nu}(\xi,t) =  \frac{- e^2}{(P+P')^+}\, \bar{u}(P^{\prime}) 
\Big[\,
   g_T^{\mu\nu} \, \Big(
      {\mathcal{H}(\xi,t)} \, \gamma^+ +
      {\mathcal{E}(\xi,t)} \, \frac{i \sigma^{+\rho}\Delta_{\rho}}{2 M}
   \Big) 
 %  \nonumber\\ &&\phantom{AAAAAAAAaaAA}
   +i\epsilon_T^{\mu\nu}\, \Big(
    {\widetilde{\mathcal{H}}(\xi,t)} \, \gamma^+\gamma_5 +
      {\widetilde{\mathcal{E}}(\xi,t)} \, \frac{\Delta^{+}\gamma_5}{2 M}
    \Big)
\,\Big]\, u(P) \, ,
\label{eq:amplCFF}
\end{equation}
\begin{figure}[h]
\begin{center}
\includegraphics[width= 0.4\textwidth]{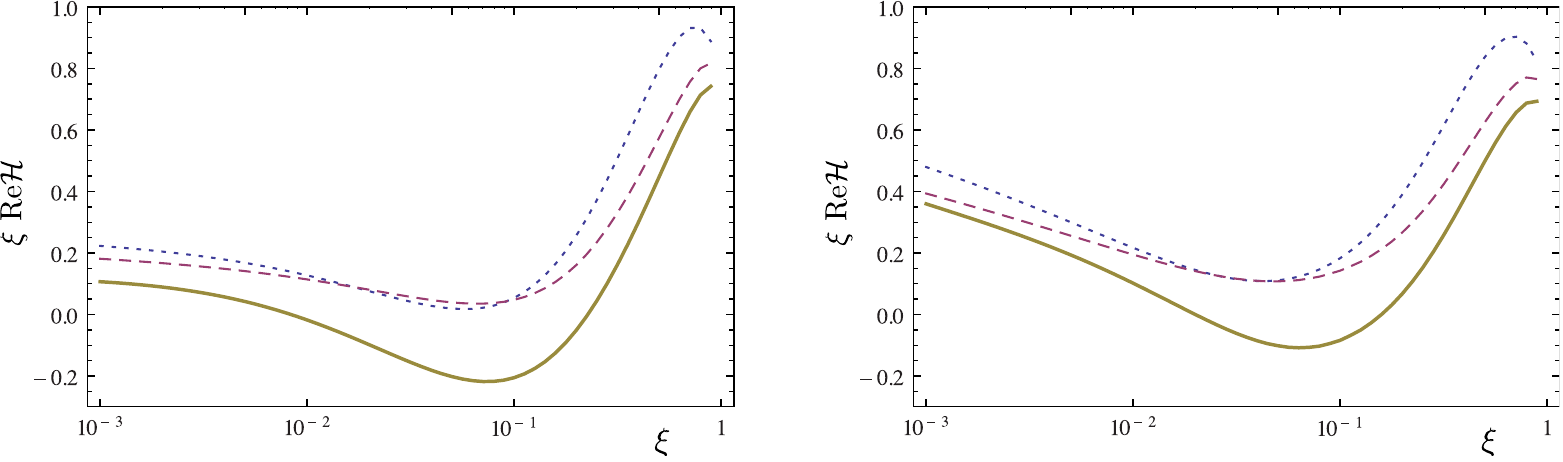} ~~~~~~  \includegraphics[width= 0.4\textwidth]{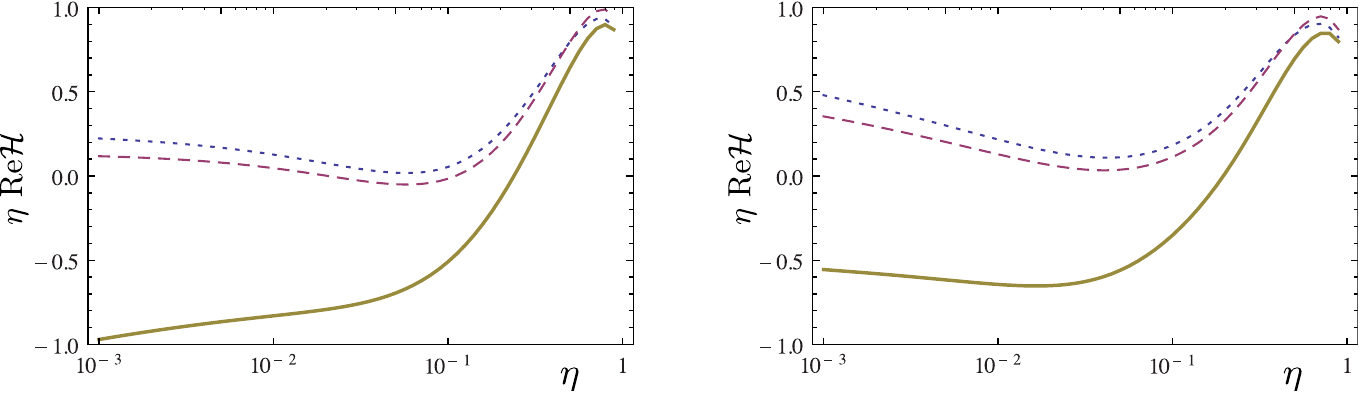} 
  \includegraphics[width= 0.4\textwidth]{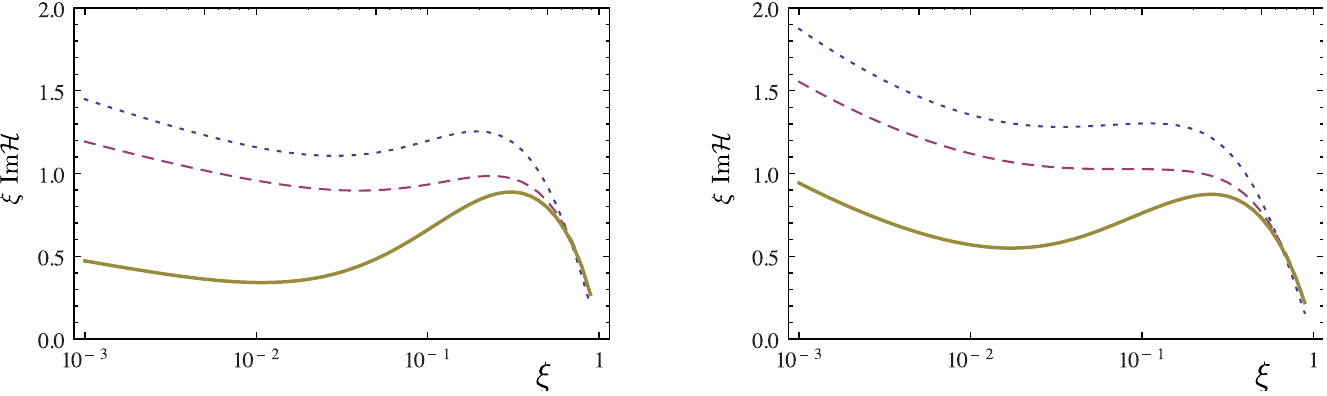}  ~~~~~~  \includegraphics[width= 0.4\textwidth]{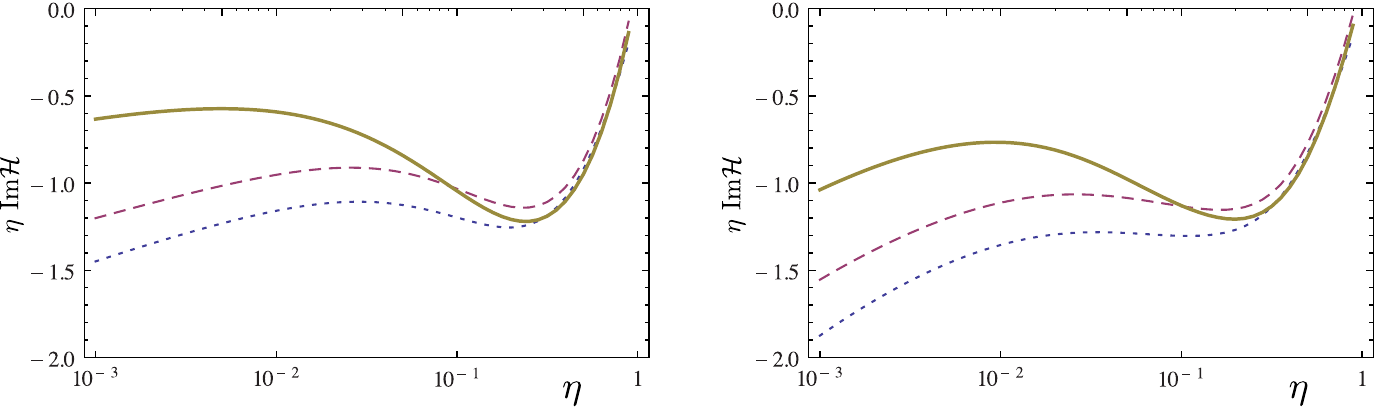} 
\caption{The real and imaginary parts of the {\it spacelike} $\mathcal{H}(\xi)$ (left) and {\it timelike} $\mathcal{H}(\eta)$ (right) Compton Form Factor multiplied by $\xi$, as a function of $\xi$ (or $\eta$) in the double distribution model based on GK (column 1 and 3) and MSTW08 (column 2 and 4) parametrizations, for $
\mu_F^2=Q^2=4 \gev^2$ and $t= -0.1 \gev^2$. In all plots, the LO result is shown as the dotted line, the full NLO result by the solid line and the NLO result without the gluonic contribution as the dashed line.}
\label{fig:DVCSRe2x2}
\end{center}
\end{figure}
with the CFFs defined, for instance in the case of $\mathcal{H}(\xi,t)$, as :
\begin{eqnarray}
\mathcal{H}(\xi,t) &=& + \int_{-1}^1 dx \,
\left(\sum_q T^q(x,\xi)H^q(x,\xi,t)
 + T^g(x,\xi)H^g(x,\xi,t)\right) \; . 
% \nonumber \\
%\widetilde {\mathcal{H}}(\xi,t) &=& - \int_{-1}^1 dx \,
%\left(\sum_q \widetilde {T}^q(x,\xi)\widetilde {H}^q(x,\xi,t) 
%+\widetilde {T}^g(x,\xi)\widetilde {H}^g(x,\xi,t)\right).
\label{eq:CFF}
\end{eqnarray}
%%%%%%%%%%%%%%%%%%%%%%%%%%%%%%%%%%%%%%%%%%%%%%%%%%%%%%%%%%%%%%%%%%%%%%%%%%%%%%%
\section{Gluonic effects to Compton form factors and to DVCS / TCS observables}

%%%%%%%%%%%%%%%%%%%%%%%%%%%%%%%%%%%%%%%%%%%%%%%%%%%%%%%%%%%%%%%%%%%%%%%%%%%%%%%%
To estimate Compton Form Factors (CFF), we use  the  NLO calculations of the coefficient functions which have been calculated in the DVCS case in the early days of GPD studies and more recently for  the TCS case \cite{PSW2}, the two results being simply related thanks to the analyticity (in $Q^2$) properties of the amplitude \cite{Muller:2012yq}:
%%%%%%%%%%%%%%%%%%%%%%%%%%%%%%%%%%%%%%%%%%%%%%%%%%%%%%%%%%%%%%%%%%%%%%%%%%%%%%%%
\begin{eqnarray}
^{TCS}T(x,\eta) = \pm \left(^{DVCS}T(x,\xi=\eta) +  i \pi C_{coll}(x,\xi = \eta)\right)^* \,,
\label{eq:TCSvsDVCS}
\end{eqnarray}
where $+$~$(-)$ sign corresponds to vector (axial) case.
Using two GPD models based on Double Distributions (DDs), as discussed in detail in Ref.  \cite{Moutarde:2013qs} : the Goloskokov-Kroll (GK) model \cite{GK3} and a model based on the MSTW08 PDF parametrization \cite{Martin:2009iq}, we get the results shown in Fig. 1 for the real and imaginary parts of the spacelike and timelike dominant CFF $\mathcal{H}(\xi,t) $ and $\mathcal{H}(\eta,t) $.  Comparing dashed and solid lines in the upper panels, one sees that gluonic contributions are so important that they even change the sign of the real part of the CFF, and are dominant for almost all values of the skewness parameter. A milder conclusion arises from a similar comparison of the lower panels; the gluonic contribution to  the imaginary part of the CFF remains sizeable for values of the skewness parameter up to $0.3$. 
%%%%%%%%%%%%%%%%%%%%%%%%%%%%%%%%%%%%%%%%%%%%%%%%%%%%%%%%%%%%%%%%%%%%%%%%%%%%%%%%
%%%%%%%%%%%%%%%%%%%%%%%%%%%%%%%%%%%%%%%%%%%%%%%%%%%%%%%%%%%%%%%%%%%%%%%%%%%%%%%%
%\section{Gluonic effects on some DVCS and TCS observables}
\begin{figure}[t]
\begin{center}
  \includegraphics[width=7cm]{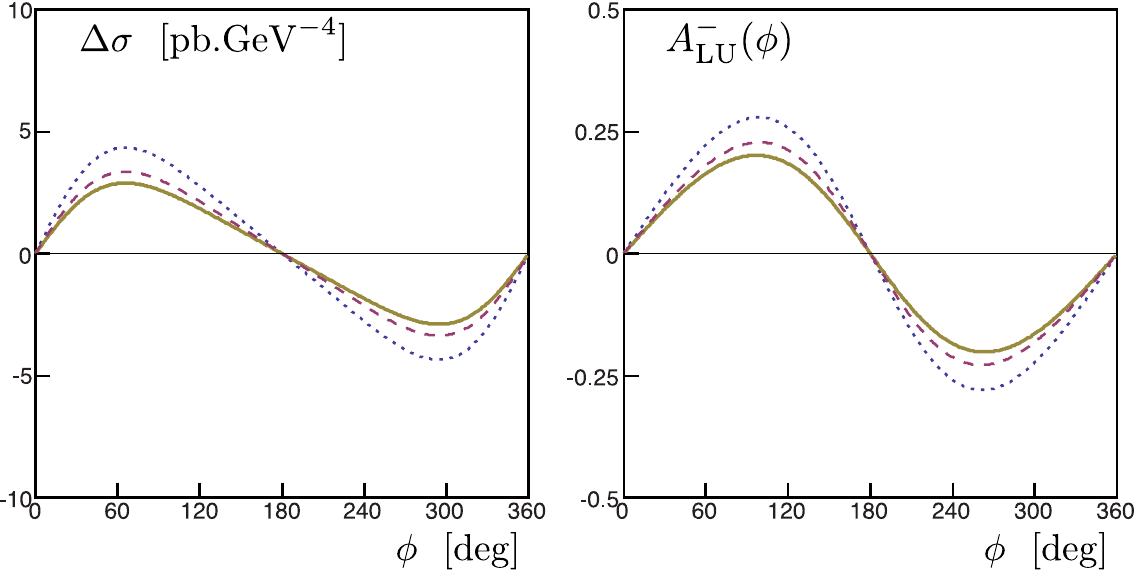} 
%   \includegraphics[width=8cm]{jlab_modshort.pdf}  
%\caption{From left to right, the total DVCS cross section in pb/GeV$^4$, the difference of cross sections for opposite lepton helicities in pb/GeV$^4$, the corresponding asymmetry, all as a function of the usual $\phi $ angle (in the Trento convention \cite{Bacchetta:2004jz})  for $E_e=11 \gev, \mu_F^2=Q^2=4$~GeV$^2$ and $t= -0.2$~GeV$^2$. The GPD $H(x,\xi,t)$ is parametrized by the GK model. The contributions from other GPDs are not included. In all plots, the LO result is shown as the dotted line, the full NLO result by the solid line and the NLO result without the gluonic contribution as the dashed line. The Bethe-Heitler contribution appears as the dash-dotted line in the cross section plots (left part).}
\caption{The difference of DVCS cross sections for opposite lepton helicities in pb/GeV$^4$ (left) and the corresponding asymmetry (right),  as a function of $\phi $    for $E_e=11 \gev, \mu_F^2=Q^2=4$~GeV$^2$ and $t= -0.2$~GeV$^2$. The GPD $H(x,\xi,t)$ is parametrized by the GK model. The contributions from other GPDs are not included.  }
\label{fig:c1}
\end{center}
\end{figure}

%%%%%%%%%%%%%%%%%%%%%%%%%%%%%%%%%%%%%%%%%%%%%%%%%%%%%%%%%%%%%%%%%%%%%%%%%%%%%%%%

\begin{figure}[hb!]
\begin{center}
  \includegraphics[width=10 cm]{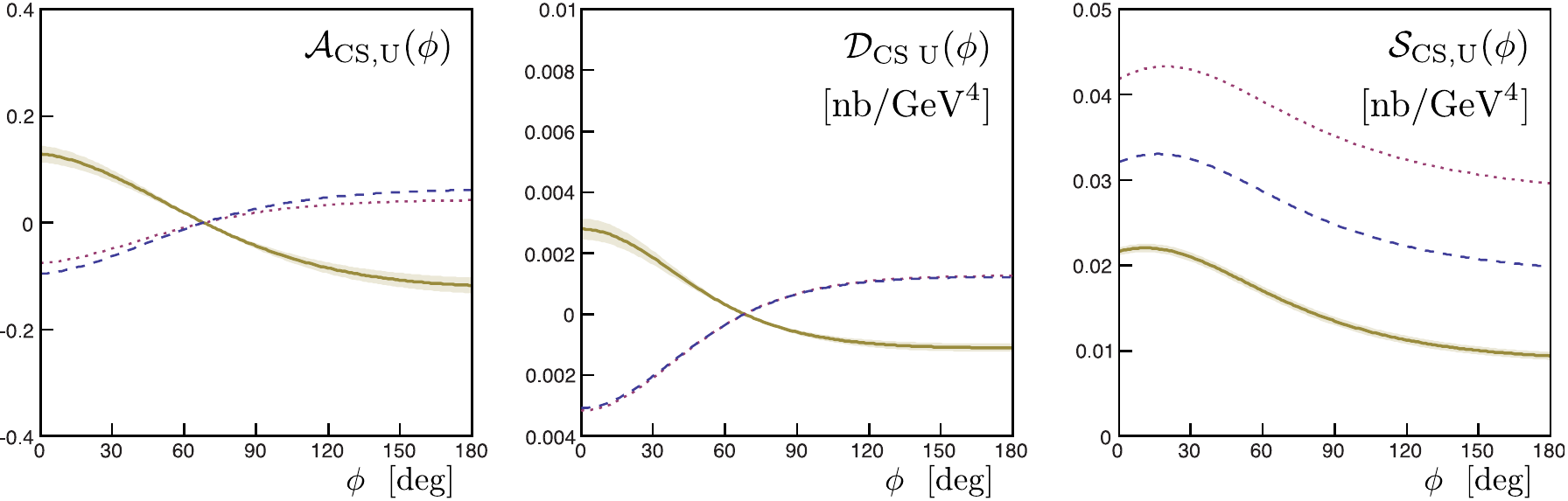}  
\caption{The DVCS observables for the COMPASS experiment, from left to right, mixed charge-spin asymmetry, mixed charge-spin difference and mixed charge-spin sum (in nb/GeV$^4$). The kinematical point is chosen as $ \xi=0.05, Q^2=4$~GeV$^2$, $t=-0.2$~GeV$^2$.  The GPD $H(x,\xi,t)$ is parametrized in the double distribution model based on the MSTW08 parametrization. The contributions from other GPDs are not included.}
\label{fig:compass}
\end{center}
\end{figure}
%%%%%%%%%%%%%%%%%%%%%%%%%%%%%%%%%%%%%%%%%%%%%%%%%%%%%%%%%%%%%%%%%%%%%%%%%%%%%%%%
\begin{figure}[t!]
\begin{center}
  \includegraphics[width=5 cm]{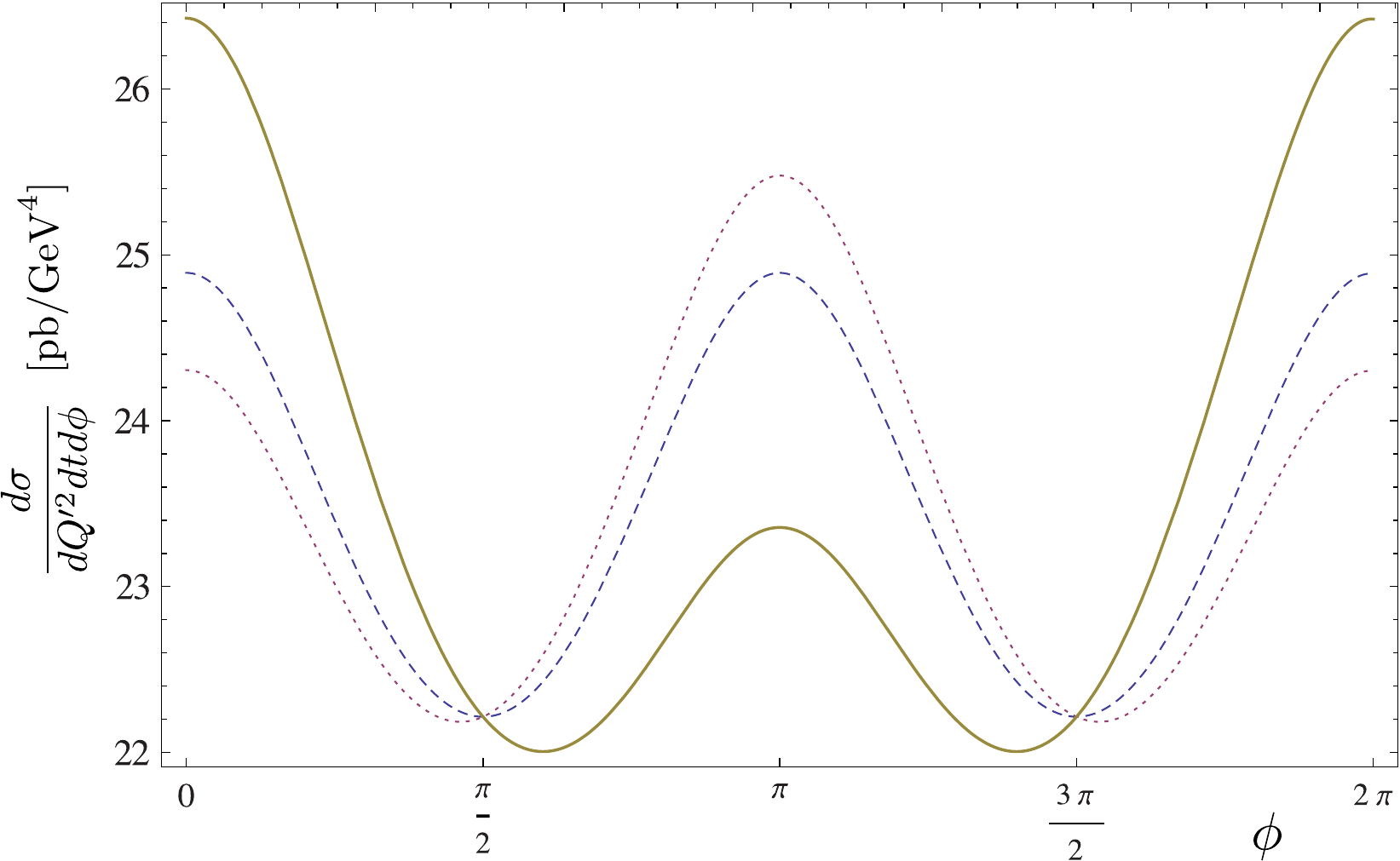}
  \hspace{1cm}
   \includegraphics[width=0.33\textwidth]{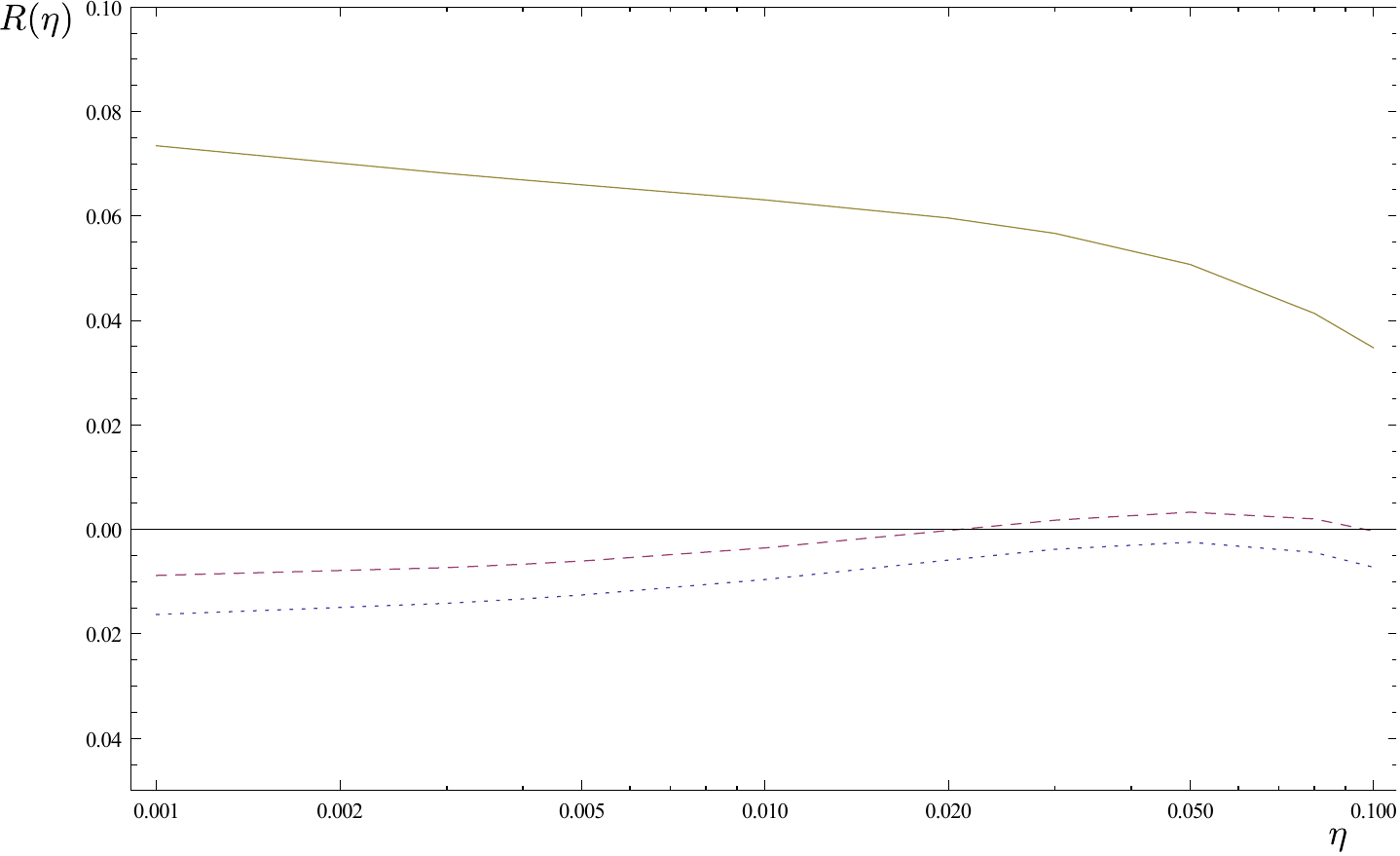} ~~~
\caption{(Left) The $\phi$ dependence of the cross-section at $E_\gamma = 10$ GeV,  $Q^2 =  \mu ^2 = 4$~GeV$^2$,  and $t= -0.1$~GeV$^2$ integrated over $\theta \in (\pi/4,3\pi/4)$: pure Bethe-Heitler contribution (dashed), Bethe-Heitler plus interference contribution at LO (dotted) and NLO (solid). (Right) The ratio R defined by Eq. (\ref{eq:Rratio}) 
as a function of $\eta$. }
\label{fig:xsec_phidep}
\end{center}
\end{figure}

%%%%%%%%%%%%%%%%%%%%%%%%%%%%%%%%%%%%%%%%%%%%%%%%%%%%%%%%%%%%%%%%%%%%%%%%%%%%%%%%

%%%%%%%%%%%%%%%%%%%%%%%%%%%%%%%%%%%%%%%%%%%%%%%%%%%%%%%%%%%%%%%%%%%%%%%%%%%%%%%%
Let us now show the effects of the gluonic contributions to some of the DVCS and TCS observables at moderate energies. Fig. \ref{fig:c1} shows the difference and asymmetry for the lepton helicity dependent observables measured at JLab. The difference between the dotted and solid lines demonstrates that NLO contributions are important, whereas the difference between the dashed and solid lines shows that gluon contributions should not be forgotten even at low energy when a precise data set is analyzed. Fig. \ref{fig:compass}, which shows mixed charge spin observables  in COMPASS kinematics, magnifies the importance of gluonic contributions.
%%%%%%%%%%%%%%%%%%%%%%%%%%%%%%%%%%%%%%%%%%%%%%%%%%%%%%%%%%%%%%%%%%%%%%%%%%%%%%%%

%\begin{figure}[b!]
%\begin{center}
%%\epsfxsize=0.8\textwidth
%  \includegraphics[width=0.33\textwidth]{Asymmetry_E_10_t_01_Q_2_nice.pdf}
%%   \includegraphics[width=0.4\textwidth]{Asymmetry_E_10_t_01_Q_2_nice.pdf}
%  \hspace{0.05\textwidth}
%  \includegraphics[width=0.33\textwidth]{Asymmetry_xi_mod.pdf}
% %   \includegraphics[width=0.4\textwidth]{Asymmetry_xi_nice.pdf}
%\caption{
%(Left) Photon beam  circular polarization asymmetry as a function of $\phi$, for $t=-0.1$~GeV$^2$, $Q^2 =  \mu^2 = 4$~GeV$^2$, integrated over $\theta \in (\pi/4,3\pi/4)$ and for $E_\gamma = 10$~GeV ($\eta \approx 0.11$).
%(Right) The $\eta$ dependence of  the photon beam  circular polarization asymmetry for  $Q^2 =  \mu^2 = 4$~GeV$^2$,  and $t= -0.2$~GeV$^2$ integrated over $\theta \in (\pi/4,3\pi/4)$. The LO result is shown as the dotted line, the full NLO result by the solid line and the NLO result without the gluonic contribution as the dashed line.}
%\label{fig:Asymmetry_xi}
%\end{center}
%\end{figure}

%%%%%%%%%%%%%%%%%%%%%%%%%%%%%%%%%%%%%%%%%%%%%%%%%%%%%%%%%%%%%%%%%%%%%%%%%%%%%%%%

With respect to TCS, since the integrated cross section is dominated by the Bethe-Heitler process at low and medium energies, one needs to analyze differential observables to get information on the TCS amplitude. We show on Fig.~\ref{fig:xsec_phidep}a the azymuthal dependence for JLab kinematics; the difference between the dotted and solid lines demonstrates the importance of NLO effects on the TCS amplitude. On Fig.~\ref{fig:xsec_phidep}b we plot the ratio R, defined by
\begin{equation}
R(\eta) =  \frac{2\int_0^{2 \pi} d \varphi \,\cos \varphi\, \frac{dS}{dQ^2dtd\varphi}}{\int_0^{2 \pi} d \varphi\frac{dS}{dQ^2dtd\varphi}}
\,,
\label{eq:Rratio}
\end{equation}
where $S$ is a weighted cross section given by Eq.~(43) of Ref.~\cite{Berger:2001xd}. It is plotted as a function of the skewness $\eta$  for  $Q^2 =  \mu^2 = 4$~GeV$^2$,  and $t= -0.2$~GeV$^2$. In the leading twist the numerator is linear in the real part of the CFFs, and the denominator, for the kinematics we consider, is dominated by the Bethe - Heitler contribution. The inclusion of NLO corrections to the TCS amplitude is indeed dramatic for such an observable and induces also a change of sign. As shown on Fig.~\ref{fig:xsec_phidep}b those corrections are dominated by the gluon contribution.

In conclusion, let us stress that our results point to the importance of understanding higher order effects, and maybe of resumming series of large contributions \cite{Altinoluk:2012nt}. On the basis of the related study  \cite{Anikin:2004jb} on scale fixing procedure in exclusive meson electroproduction, we do not expect that these gluonic contributions can be significantly reduced by a judicious choice of the factorization scale, simultaneously in the timelike and spacelike cases, and for both the real and imaginary parts of the Compton form factors.
This indicates that current attempts  \cite{Guidal:2009aa,Kroll:2012sm, Guidal:2013rya,  Kumericki:2009uq, Kumericki:2013br} to extract physics from DVCS data are but a first step in a long range program which will include data analysis from low to very high energy \cite{EIC}.

%%%%%%%%%%%%%%%%%%%%%%%%%%%%%%%%%%%%%%%%%%%%%%%%%%%%%%%%%%%%%%%%%%%%%%%%%%%%%%%%

{
\small
\section*{Acknowledgements} 
This work is partly supported by the Polish Grants NCN No DEC-2011/01/D/ST2/02069, by the Joint Research Activity "Study of Strongly Interacting Matter" (acronym HadronPhysics3, Grant Agreement n.283286) under the Seventh Framework Programme of the European Community, by the GDR 3034 PH-QCD,  the ANR-12-MONU-0008-01 and  the COPIN-IN2P3 Agreement.
}

%
% BibTeX or Biber users please use (the style is already called in the class, ensure that the "woc.bst" style is in your local directory)
% \bibliography{name or your bibliography database}

\end{document}